# Magnetic Dynamics of Iron-Oxide Nanoparticles in Frozen Ferrofluids and Ferronematics

G. F. Goya[1], S. L. Gómez[2] and S. M. Shibli[2]

[1] Magnetic Materials Laboratory, Physics Institute, University of São Paulo, São Paulo, Brazil.

[2] Complex Fluids Group, Physics Institute, University of São Paulo, São Paulo, Brazil.

**Abstract.** We present a detailed study of the magnetic behavior of iron-oxide ($\gamma$-$Fe_2O_3$ and $Fe_3O_4$) nanoparticles constituents of ferrofluids (FF's) with average particle sizes $<d>$ = 2.5 and 10 nm. The particles were dispersed in the frozen liquid carrier (pure FF) and in a frozen lyotropic liquid crystalline matrix in the nematic phase or ferronematic (FN) (ferrolyomesophase). Both FF and FN phases displayed superparamagnetic (SPM) behaviour at room temperature, with blocking temperatures $T_B$ ~ 10 and 100 K for $<d>$ = 2.5 and 10 nm, respectively. Dynamic ac susceptibility measurements showed a thermally activated Néel-Brown dependence of the blocking temperature with applied frequency. Our results show that dipolar interactions are small, but non-negligible, as compared to the single-particle energy barriers from magnetic anisotropy. From the fit of ac susceptibility we calculated the effective magnetic anisotropy constant $K_{eff}$ for 2.5 nm maghemite particles. Although interparticle interactions present in highly diluted samples do not appreciably modify the dynamic magnetic behavior of isolated particles, the calculated magnetic anisotropy were abut one order of magnitude larger that the bulk materials, suggesting the existence of large surface anisotropy. Using the thermally activated model to fit the dynamic data yielded effective energy barriers $E_a$ = $3.5 \times 10^{-21}$ J. From these data, we obtained $K_{eff}$ = 422 kJ/m$^3$ for the single-particle effective magnetic anisotropy.

**Introduction.**

Nanostructures, a size regime between molecular and submicron-size structures, possess high surface-to-volume ratios, yielding novel properties. Many biological and biomedical applications uses such nanoscopic systems in a magnetic fluid or *ferrofluid*, a colloidal suspension of magnetic oxide nanograins dispersed in a liquid carrier. Depending on the characteristic of the liquid carrier (polar or apolar character and pH), magnetic grains are coated with an appropriate agent for stability purposes. The synthesis and applications of magnetic fluids for biomedical applications have passed the speculative stage of development, to enter into the well-based standard laboratory procedures. Nowadays, these particles are being used in magnetic resonance contrast enhancement, cell and DNA separation, drug delivery, and gene cloning. Often, the desired properties include large magnetic moments per formula unit, making iron oxide particles such as $Fe_3O_4$ and $\gamma$-$Fe_2O_3$ appealing for these applications. For biocompatible purposes, particles are usually coated with noble metals (Gd, Pt, Ag), or organic substances such as dextran. [1] In spite of the above advancements on bio-magnetic applications, there is still a lack of precise knowledge on the mechanisms linking particle shape, size distribution and surface structure, to the resulting magnetic properties of the carrier magnetic nanoparticles.

---

[1] Corresponding author. E-mail: goya@if.usp.br

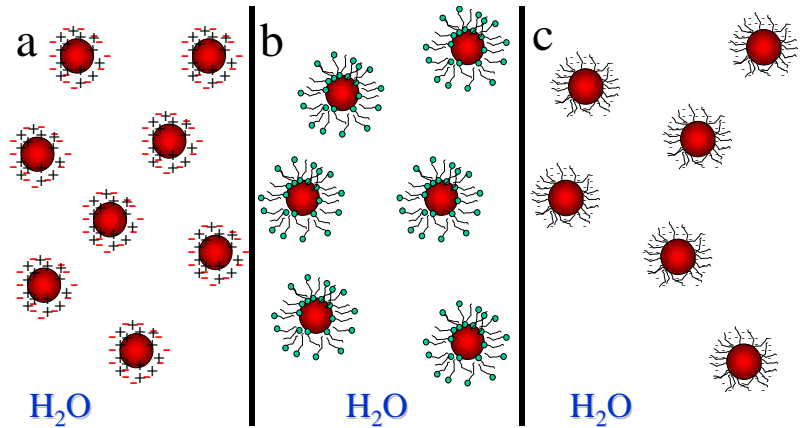

**Figure 1. Illustrative picture of different coatings for stabilize the magnetic grains
in a particular liquid carrier: Ionic (a), surfacted (b), and citrated (c) types.**

The lyotropic liquid crystals as those used in this work are organized molecular systems formed by a mixture of amphiphilic molecules (molecules with both hydrophobic and hydrophilic parts) and a solvent, most often water, and their structure depends on temperature and relative composition of the constituents [2]. Amphiphilic aggregates in the nematic mesophase, named *micelles* (typical dimensions 10 nm), have a bilayer structure, resembling cell membranes.

Regarding the magnetic grains, both maghemite and magnetite have a cubic spinel structure with two interacting magnetic sublattices. The strong superexchange coupling between A and B magnetic sublattices yields antiparallel alignment, and thus ferrimagnetic order below $T_C$. The first-order magnetocrystalline anisotropy constants are (at room temperature) $K_1$ = 13.5 kJ/m$^3$ and 4.7 kJ/m$^3$ for $Fe_3O_4$ and $\gamma$-$Fe_2O_3$, respectively. Interparticle interactions can modify the particle energy barriers, yielding a shift in the observed blocking temperature [3] due to collective freezing of the magnetic moments. Concurrently with dipolar interactions, other sources of magnetic anisotropy (e.g. surface effects) can have a major impact on the blocking process. In this scenario the freezing process, analyzed as a superspin-glass transition, is subject of intense debate. In this work, we present a magnetic characterization of magnetite ($Fe_3O_4$) and maghemite ($\gamma$-$Fe_2O_3$) nanoparticles with mean size <d> of 10nm and 2.5 nm, respectively, with different concentrations to analyze the effects of dipolar interactions on the magnetic dynamic of the constituent particles.

**Experimental Procedure**

We have studied the magnetic interaction between the nanograins in frozen samples of a pure FF and a FF-doped lyotropic liquid crystals in the nematic phase. The magnetic particles of the starting FF were coated with a surfactant (figure 1) to prevent agglomeration, and suspended in an nonmagnetic solvent as illustrated in figure 1. Ferrofluids are magnetic colloids in which magnetic oxide nanograins are dispersed in a liquid carrier. The two main types used in this work were surfacted (SFF) and citrated (CFF) ferrofluids (see fig. 1). Magnetite ($Fe_3O_4$) particles having <d> = 10 nm were studied from a commercial SFF from Ferrotec® Company, whereas the <d> = 2.5 nm maghemite ($\gamma$-$Fe_2O_3$) particles were from a non-commercial CFF. The ferronematic (FN) phases were obtained by mixing the corresponding FF with lyotropic crystals with different concentrations. The relative volume concentration of the samples, $R_V = 100 \times \frac{\text{vol. magnetic phase}}{\text{total volume}}$ is shown in Table I.

Static and dynamic magnetic measurements as a function of frequency and temperature were performed in a commercial SQUID magnetometer (Quantum Design). Zero-field-cooled (ZFC) and field-cooled (FC) curves were taken between 5K and 250 K, for different values of cooling field $H_{FC}$ (10 Oe < H < 70 kOe). AC susceptibility of the samples was obtained for frequencies from 10$^{-2}$ to $10^3$ Hz, as a function of temperature. Magnetization data were obtained by first cooling the sample from room temperature in zero applied field (ZFC process) to the basal temperature. Then a field was applied and the variation of magnetization was measured with increasing temperature up to T = 250 K. After the last point was measured, the sample was cooled again to the basal temperature keeping the same field (FC process); then the M *vs.* T data was measured for increasing temperatures.

| Label | Sample | $<d>$ (Å) | $R_V$% (vol.) | $T_B$ (K) | $H_C$(5K) (Oe) | $H_C$(30K) (Oe) | $M_R/M_S$ (5K) | $M_R/M_S$ (30K) |
|---|---|---|---|---|---|---|---|---|
| SFF0 | Commercial Ferrofluid (SFF) | 100 | 1.8 | 218 | 80(5) | 70(5) | 0.12(2) | 0.12(2) |
| SFFC | Concentrated SFF Ferronematic | 100 | ~$10^{-2}$ | 228 | 125(10) | 103(5) | -- | -- |
| SFFD | Diluted SFF Ferronematic | 100 | ~$2 \times 10^{-4}$ | 169 | 100(10) | 98(8) | -- | -- |
| CFF0 | Non-commercial Ferrofluid (CFF) | 25 | 1.1 | 9.1 | 55 (5) | 12(5) | 0.06(1) | 0.00(1) |
| CFFC | Concentrated CFF Ferronematic | 25 | 0.257 | 13.9 | 115(12) | 5(5) | -- | -- |
| CFFD | Diluted CFF Ferronematic | 25 | 0.00076 | 11.1 | 70(10) | 5(5) | -- | -- |
| CFFY | Dried CFF Ferrofluid | 25 | Unknown | 11.8 | 65(5) | 13(5) | 0.07(1) | 0.01(1) |

**Table 1. Description, structural and magnetic parameters of the samples studied in this work: mean particle diameter $<d>$, volumetric concentration $R_V$ (%), Blocking temperature $T_B$ (at H = 100 Oe), coercive fields $H_C$ and remanence to saturation ratio $M_R/M_S$.**

**Results and Discussion**

The structural and magnetic parameters of the studied samples are condensed in Table I. The different average particle diameter for the two SFF and CFF samples are reflected in the blocking temperatures $T_B$ (fig. 2). For the SFF samples, the ZFC magnetization curves showed a broad maximum at ~200 K, which extends above the freezing temperature of the matrix (~260 K). Therefore, at our maximum measuring temperatures of ca. 250 K, some of the particles might be still in the blocked state. For the CFF samples, the $T_B$ values vary between 9(1) and 14(1) K, depending on the particle concentration. Figure 2 clearly shows that, for these samples, there will be no significant blocked particles at ~250 K. Indeed, the coercive fields at T = 30 K for SFF and CFF samples show this difference (fig. 2), since samples CFF have $H_C = 0$ but a significant $H_C$ is observed for SFF samples, due to the blocked fraction. Interestingly, the blocking temperatures displayed in Table I show a measurable increase from the FF phase to the FN concentrated and diluted. We have checked the effect of interparticle interaction by evaporating a large fraction of the solvent in the CFF sample, obtaining a 'dried' sample (CFFY) where particles are closer to each other. It can be observed that for this sample, $T_B$ shifts upwards by ~30% of the CFF0 sample, indicating that the effect of dipolar interactions goes in this direction. Since the average interparticle distance in FN phases is much larger (high dilution) we attribute the observed increase in $T_B$ to particle agglomeration formed in these samples. As the concentration used were well above the standards for colloidal stability, this effect is likely to occur.

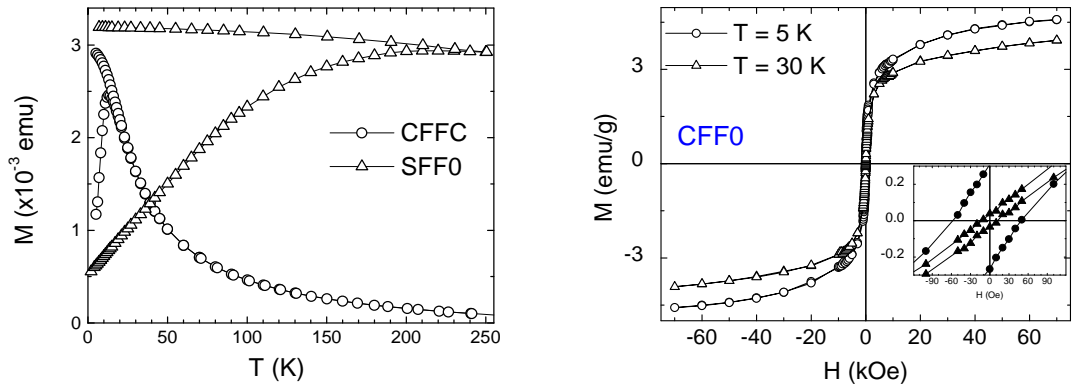

**Figure 2.** Left panel: AFC/FC curves for surfacted (<d> = 10 nm) and citrated (<d> = 2.5 nm) ferrofluids, showing their respective blocking temperatures. Right panel: Typical magnetization curves as a function of applied field.

For each set of samples (CFF and SFF), a slight increase in $H_C$ can be observed from the pure ferrofluid to the ferronematic phases. Table I also display the remanence to saturation $M_R/M_S$ ratio, except for the FN samples, where this ratio has not clear meaning due to the diamagnetic contribution of the matrix, which dominates at high fields. However, for the measured ratios the observed value is much smaller than the expected. At T = 5 K, the $M_R/M_S$ values for the samples studied (see Table I) are much smaller than the expected R = 0.5 value for noninteracting, randomly oriented particles with uniaxial symmetry. Deviations from the non-interacting values are expected, and moreover these low values have been related to the existence of inter-particle interactions of antiferromagnetic nature. [4]

We have investigated the magnetic dynamics of the <d> = 2.5 nm ferrofluid samples, in order to detect the effect of interparticle interactions on the single-particle behaviour. Figure 3 displays the temperature dependence of $\chi'(T)$ and $\chi''(T)$ of the smallest particles (samples CFF0 and CFFY) for different frequencies $f$ ranging from 20 mHz to 1 kHz. The data for both components $\chi'(T)$ and $\chi''(T)$ exhibit the expected behavior of SPM systems, i.e., the occurrence of a maximum at a temperature $T_m$ (~$T_B$) for both $\chi'(T)$ and $\chi''(T)$ components, which shifts towards higher values with increasing frequency. [3] From the empirical relationship $\Phi = \Delta T_B/T_B[\Delta\log_{10}(f)]$, where $\Delta T_B$ is the difference between $T_B$ measured in the $\Delta\log_{10}(f)$ frequency interval, we obtained $\Phi = 0.089(2)$ and $0.067(2)$ for CFF0 and CFFY, respectively, indicating the presence of interparticle interactions the usually yield $\Phi$ values lower than the 0.1-0.13 theoretically expected. [5,6] Accordingly, the effect of dipolar interactions in the more concentrated sample (CFFY) is reflected in the lower $\Phi$ value. It is well known that the dynamic response of an ensemble of fine particles is determined by the measuring time $\tau_m$ (or frequency) of each experimental technique. As the reversion of the magnetic moment in a single-domain particle over the anisotropy energy barrier $E_a$ is assisted by thermal phonons, the relaxation time $\tau$ exhibits an exponential dependence on temperature characterized by

$$\tau = \tau_0 \exp\left(\frac{E_a}{k_B T}\right) \qquad (1)$$

where $\tau_0$ is in the $10^{-9}$ - $10^{-11}$ s range for SPM systems. In the absence of an external magnetic field, the energy barrier can be assumed to be $E_a = K_{eff} V$, where the proportionality constant $K_{eff}$ is an effective magnetic anisotropy.

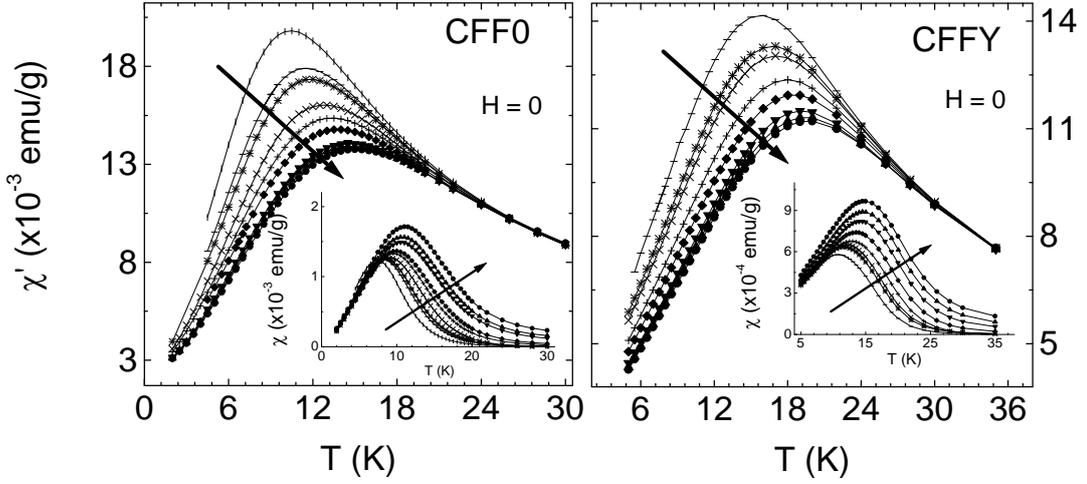

**Figure 3.** Temperature dependence of the in-phase (real) component $\chi'(T)$ of the magnetic susceptibility for CFF samples, at different excitation frequencies, taken in zero external magnetic field. The arrows indicate increasing frequencies. Inset: Out of phase (imaginary) component $\chi''(T)$.

The linear dependence of $\ln(f)$ versus $1/T_B$ observed in fig. 4 indicates that the Néel-Brown model correctly suits the behavior of CFF samples. From the fitting of the experimental $T(f,H=0)$ data using Eq. 1, we obtained the values of $E_a/k_B = 250(7)$ K for CFF0 sample and $E_a/k_B = 412(11)$ K for the more concentrated sample CFFY. This increment in $E_a$ agrees with the expected contribution from dipolar interactions to the single-particle effective energy barrier. Using the average particle diameter $<d> = 2.5$ nm, we obtained $K_{eff} = 422$ kJ/m$^3$ and 695 kJ/m$^3$ for the effective anisotropies for samples CFF0 and CFFY, respectively.

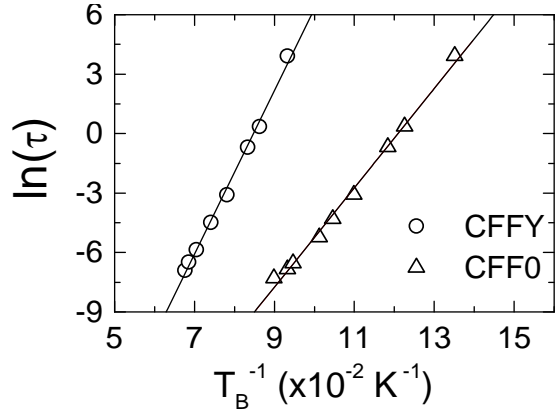

**Figure 4.** Arrhenius plot of the relaxation time $\tau$ vs. $T_B^{-1}$ obtained from the imaginary component $\chi''(T)$ of figure 3. Solid line is the best fit using Eq. (1).

The larger $E_a$ value found in concentrated sample should be attributed to the increase of dipolar interactions that add to the single particle energy barrier. However, the $K_{eff}$ value calculated for the most diluted sample is approximately two orders of magnitude larger that the magnetocrystalline anisotropy constant of bulk maghemite $K_1^{bulk} = 4.7$ kJ/m$^3$, indicating the existence of additional anisotropy sources. In previous works on more concentrated nanoparticle systems, large values of particle anisotropy are usually related to surface effects and/or the effect of dipolar interactions [3]. The contribution from particle-particle interactions to the experimental data is difficult to disentangle from the single-particle magnetic anisotropy. In addition to the intrinsic sources of particle anisotropy (e.g. shape, stress and magnetocrystalline anisotropies), interparticle interactions (mainly the long-range dipolar interaction) can also modify the energy barrier values. In a previous work on a ferrofluid consisting in $Fe_3O_4$ particles of d = 5nm,[7] it was found that dipolar interactions were already noticeable for concentrations $R_V$ of ~2 % vol. of magnetic particles. However, the high dilution of our samples (except for the dried sample, see Table 1) guarantees that the value of $K_{eff}$ obtained should be essentially the single-particle value. Surface effects have been invoked as a large source of magnetic anisotropy in Cobalt [8] and maghemite [9] nanoparticles. Our data clearly indicates that both the similarity between $K_{eff}$ for liquid and dried samples and the large $K_{eff}$ value are a consequence of an intrinsic property of these maghemite particles, suggesting that surface anisotropy is dominant.

**Summary**

We have studied the effect of interparticle interactions on the magnetic properties of single-domain $Fe_3O_4$ and $\gamma$-$Fe_2O_3$ nanoparticles with identical size distribution but different concentration. We found that stronger interparticle interactions yield an increase of $T_B$, as observed from agglomeration of particles in FN phases. For diluted ferrofluid samples with maghemite particles having <d> = 2.5 nm, the calculated anisotropy constant is much larger that the bulk value, indicating a large contribution to the magnetic anisotropy, which is likely to be located at the particle surface.

**Acknowledgements**

We wish to acknowledge Prof. A. Bee for providing the citrated ferrofluid used in this work. Financial support from Brazilian agencies FAPESP and CNPq is also acknowledged.

**References**

[1] K. Nishimura, M. Hasegawa, Y. Ogura, T. Nishi, K. Kataoka, H. Handa, and M. Abe, J. Appl. Phys. **91**, 8555 (2002). See also Lee, J.; Isobe, T.; Senna, M. *J. Colloid Interface Sci*. **1996**, 177, 490.

[2] Phase Transition in complex Fluids, Eds. P. Tolédano and A.M. Figueiredo Neto, World Scientific, Singapore 1998.

[4] G. Hadjipanayis, D.J. Sellmyer and B. Brandt, Phys. Rev. B **23**, 3349 (1981).

[3] J.L. Dormann, D. Fiorani and E. Tronc, in *Advances in Chemical Physics*, Ed. By I. Prigogine and S.A. Rice. Vol. XCVIII 326 (1997).

[6] G.F. Goya, F. C. Fonseca, R. F. Jardim, R. Muccillo, N.L.V. Carreño, E. Longo and E.R. Leite. J. Appl. Phys. **93**, 6531 (2003).

[5] J. A. De Toro, M. A. López de la Torre, M. A. Arranz, J. M. Riveiro, J. L. Marínez, P. Palade and G. Filoti, Phys. Rev. B **64,** 094438 (2001).

[7] W. Luo, S.R. Nagel, T.F. Rosenbaum and R.E. Rosensweig, Phys. Rev. Lett. **67**, 2721 (1991).